\newcommand{\nc}{\newcommand}
\nc{\kms}{\,{km\,s$^{-1}$}}
\nc{\sgra}{Sgr\,A}
\nc{\sgrastar}{Sgr\,$\rm {A}^{*}$}
\nc{\sgraeast}{Sgr\,A\,East}
\nc{\sgrawest}{Sgr\,A\,West}
\nc{\sgracomp}{Sgr\,A\,Complex}
\nc{\as}{\arcsec}
\nc{\HI}{H\,{\sc i}}
\nc{\HII}{H\,{\sc ii}}
\nc{\CI}{C\,{\sc i}}
\nc{\hto}{H$_{2}$O}
\nc{\htmo}{H$_{2}^{16}$O}
\nc{\htio}{H$_{2}^{18}$O}
\nc{\am}{\arcmin}
\nc{\ciso}{C$^{18}$O}
\nc{\htwo}{H$_{2}$}
\nc{\ot}{O$_{2}$}
\nc{\nht}{NH$_{3}$}
\nc{\htco}{H$_{2}$CO}
\nc{\htre}{H$_{3}$O${^+}$}
\nc{\ohs}{OH-Streamer}
\nc{\msol}{{$\mathrm{M}_{\odot} $}}
\nc{\fas}{$\farcs$}

\documentclass[bibyear]{aa}
\usepackage{graphicx}
\usepackage{txfonts}

\begin{document}

\title{{\it Odin} observations of ammonia in the \sgra\ +50 \kms\
  Cloud and Circumnuclear Disk
\thanks{{\it Odin} is a Swedish-led satellite project funded jointly
  by the Swedish National Space Board (SNSB), the Canadian Space
  Agency (CSA), the National Technology Agency of Finland (Tekes), the
  Centre National d'Etudes Spatiales (CNES), France and the European
  Space Agency (ESA). The former Space division of the Swedish Space
  Corporation, today OHB Sweden, is the prime contractor, also
  responsible for {\it Odin} operations.}}

\author{Aa.\,Sandqvist\inst{1}
   \and \AA .\,Hjalmarson\inst{2} 
   \and U.\,Frisk\inst{3}
   \and S.\,Lundin\inst{4}
   \and L.\,Nordh\inst{1}
\and M.\, Olberg\inst{5}
\and G.\, Olofsson\inst{1}
}

\institute{Stockholm Observatory, Stockholm University, AlbaNova
  University Center, SE-106 91 Stockholm, Sweden\\ \email{aage@astro.su.se}
\and Department of Earth and Space Sciences, Chalmers University of
Technology, Onsala Space Observatory, SE-439 92 Onsala, Sweden
\and Omnisys Instruments AB, Solna Strandv\"ag 78, SE-171 54 Solna,
Sweden
\and OHB Sweden, PO Box 1269, SE-164 29 Kista, Sweden
\and Onsala Space Observatory, Chalmers University of Technology,
SE-439 92 Onsala, Sweden
}

\offprints{\\ Aage Sandqvist, \email{aage@astro.su.se}}

\date{Received $<$date$>$; accepted $<$date$>$}

\abstract {The {\it Odin} satellite is now into its sixteenth year of
  operation, much surpassing its design life of two years. One of
  the sources which {\it Odin} has observed in great detail is the
  \sgracomp\ in the centre of the Milky Way.} {To study the presence
  of \nht\ in the Galactic Centre and spiral arms.} {Recently, {\it Odin} has
    made complementary observations of the 572 GHz \nht\ line towards
    the \sgra\ +50 \kms\ Cloud and Circumnuclear Disk (CND).}
  {Significant \nht\ emission has been observed in both the +50 \kms\
    Cloud and the CND. Clear \nht\ absorption has also been detected
    in many of the spiral arm features along the line of sight from
    the Sun to the core of our Galaxy.} {The very large velocity width
    (80 \kms) of the \nht\ emission associated with the shock region
    in the southwestern part of the CND may suggest a
    formation/desorption scenario similar to that of gas-phase \hto\
    in shocks/outflows.}


\titlerunning{{\it Odin} observations of \nht\ towards the \sgracomp}
\authorrunning{Aa. Sandqvist et al.}
\maketitle

\section{Introduction}

The core of the Milky Way Galaxy is a region of great complexity
containing a wide variety of physical environments. At the very
centre resides a four-million-solar-mass supermassive black hole,
whose non-thermal radio continuum signature is called
\sgrastar. Orbiting around it, at a distance of one to a few pc, with
a velocity of about 100 km/s, is a molecular
torus called the Circumnuclear Disk (CND). The CND has a mass of
$10^4$ to $10^5$ \msol\ and a gas temperature of several hundred
degrees. Beyond this, there exists a large Molecular Belt consisting
predominantly of two GMCs, called the +50 and the +20 \kms\
clouds. Both GMCs are massive, about $5 \times 10^5$ \msol, with a density
$10^4$ - $10^5$ cm$^{-3}$, gas temperature 80 - 100 K, and dust
temperature 20 - 30 K (e.g. Sandqvist et al. \cite{san08}). General
reviews of the Galactic Centre have been presented by e.g. Mezger et
al. (\cite{mez96}) and Morris \& Serabyn (\cite{mor96}), with an
up-to-date introduction to the \sgracomp\ given by Ferri\`ere
(\cite{fer12}).

The {\it Odin} satellite (Nordh et al. \cite{nor03}; Frisk et
al. \cite{fri03}) has surveyed the \sgracomp\ in a number of
different molecules. While it was unsuccessful in searching for \ot\
- a $3\sigma$ upper limit for the fractional abundance ratio of
[\ot/\htwo], averaged over a 9-arcmin region, was found to be
$X$(\ot)$ \le 1.2 \times 10^{-7}$ (Sandqvist et al. \cite{san08}) -
significant amounts of \hto, CO and \CI\ were detected in many regions
of the \sgracomp\  (Karlsson et al. \cite{kar13}). Unfortunately,
there were a number of instabilities in the 572 GHz \nht\ receiver
during those observation periods which prevented us from obtaining results for
this \nht\ transition in the core region of the Galaxy, although some results
were obtained for the spiral arm features. These instabilities were
due to a loss of phase lock for this receiver early in the
mission. However, with appropriate centering of the observing frequency and
many frequency calibrations, performed using a nearby telluric
ozone line, it is possible to correct for this lack of phase
lock. This short paper now reports on new successful {\it Odin}
\nht\ observations in the \sgra\ +50 \kms\ Cloud, and in the
southwestern region of the CND which is a complex shocked region
containing a number of interacting components (Karlsson et al. \cite{kar15}). 

\section{Observations}

The {\it Odin} observations of the o-\nht\ $(1_0 - 0_0)$ line at
572.4981 GHz towards the \sgra\ +50 \kms\ Cloud at J2000.0 $17^{\rm
  h}45^{\rm m}51\fs7, -28\degr59\arcmin09\arcsec$ were performed in
April 2015, with an ON-source total integration time of 27.4 hours. The \nht\
observations towards the southwestern (SW) lobe of the Sgr A CND at
J2000.0 $17^{\rm h}45^{\rm m}39\fs7, -29\degr01\arcmin18\arcsec$  were
performed in April 2016, with an ON-source total integration time of
26.7 hours. The system temperatures were 3300 and 3400 K,
respectively. The halfpower beamwidth of {\it Odin} at the \nht\
frequency is $2\farcm1$ and the main beam efficiency is 0.89 (Frisk et
al. \cite{fri03}). The backend spectrometer was a 1050 MHz AOS with a
channel resolution of 1 MHz. For more details of {\it Odin} Galactic Centre
observations, see Karlsson et al. (\cite{kar13}).

\section{Results and Discussion}

Our two \nht\ profiles, obtained towards the +50 \kms\ Cloud and the
CND SW, are presented in Figs. 1 and 2, where they have been smoothed
to a channel resolution of 1.6 \kms. Gaussian analysis was
  performed on both profiles by visual estimation of all three
  parameters (intensity, velocity, and velocity half width) for each
  emission and absorption component and the fit was then executed in
  an unbiased manner. The Gaussian analysis fittings are also
  presented in Figs. 1 and 2, and in Tables 1 and 2 where the
    listed uncertainties are at the $1\sigma$ level. The dominant
impression in the figures is the clear emission features at positive
velocities, which  arise in the \sgracomp\ (e.g. Sandqvist
  \cite{san89}). Most of the numerous weaker absorption features at
negative velocities are caused by the various Galactic spiral arm
components along the line of sight from the Sun to the Galactic Centre
and are identifiable by their radial velocities (Sandqvist et
al. \cite{san15}). A few of these features are near the limit of
detectability, but correspond to clearly identifiable signatures
observed in e.g. OH, \hto\ and CO lines presented by Karlsson et
al. (\cite{kar13}). The rms noise levels in the two profiles are 9 and
11 mK, respectively, so the intensities of the weak absorption
features are predominantly greater than $3\sigma$. 

The very fact that we are only observing ground-state ammonia
  absorption features at negative velocities, and see no visible
  emission, against very weak continuum background sources (only 210
  and 140 mK) implies that the excitation temperature ($T_{\rm ex}$)
  of this gas must be close to the temperature of the cosmic microwave
  background ($T_{\rm CMB} = 2.725$ K), which means that the dominant
  part of the ammonia population is residing in the lowest state. This
  also means that the absorbing gas regions must have a density
  several orders of magnitude lower than the critical density of the
  ammonia line ($5 \times 10^7$ cm$^{-3}$), which is also consistent
  with their low \htwo\ column densities (see Table 3) and low visual
  extinction. From the Gaussian analysis of the absorption features,
we can now obtain column densities, $N$(\nht), of ortho-\nht\ in
a manner similar to Karlsson et al. (\cite{kar13}):
\begin{equation}
N$(\nht)$ = 3.69 \times 10^{12} \tau_{o}$(\nht)$ \Delta V_{\rm FWHM}
\label{Nnht}
\end{equation}
where $\Delta V_{\rm FWHM}$ is the line full width at half maximum and
$\tau_{o}$ is the central optical depth of the Gaussian fitted
to the absorption feature and (for $T_{\rm ex} = T_{\rm CMB}$)
  can be calculated as 
\begin{equation}
\tau_{o} = -{\rm ln} \bigg[1+{T^*_{\rm A} \over {T_{\rm cont}}}\bigg] .
\label{tau0}
\end{equation} 
In Eq. (\ref{tau0}), $T^*_{\rm A}$ is the antenna temperature of the
absorption line intensity and is a negative number and $T_{\rm
  cont}$ is the background continuum temperature, which is 210 and 140
mK for the +50 \kms\ Cloud and the CND SW, respectively (Karlsson et
al. \cite{kar13}). The results of the Gaussian analysis of the
absorption features are summarized in Tables 1 and 2.

\begin{figure}
  \resizebox{\hsize}{!}{\rotatebox{270}{\includegraphics{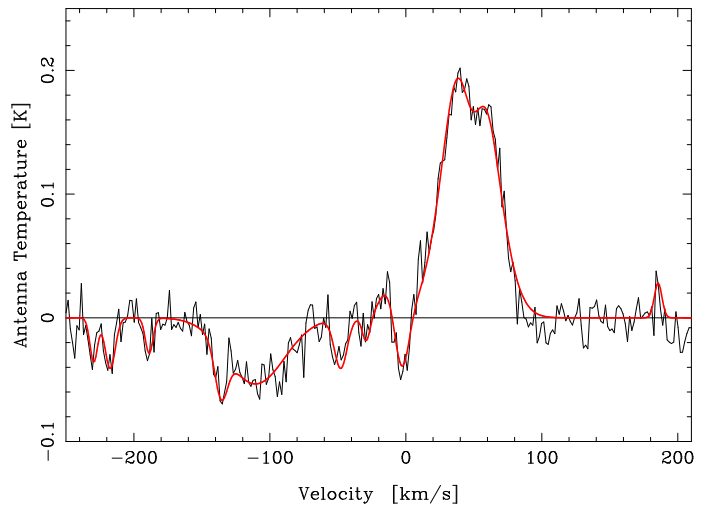}}}
  \caption{{\it Odin} observations of the 572 GHz \nht\ line towards
    the \sgra\ +50 \kms\ Cloud (black profile); the channel resolution
    is 1.6 \kms. The red profile is the result of the Gaussian analysis.}  
  \label{1}
\end{figure}
  
\begin{figure}
  \resizebox{\hsize}{!}{\rotatebox{270}{\includegraphics{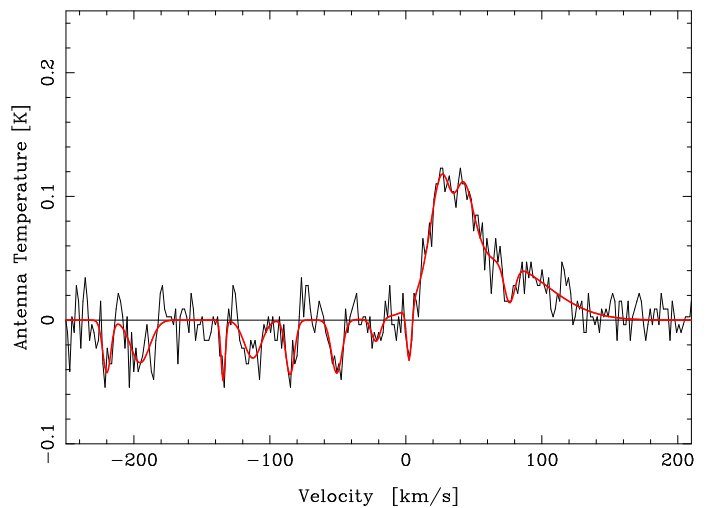}}}
  \caption{{\it Odin} observations of the 572 GHz \nht\ line towards
    the southwestern part of the \sgra\ Circumnuclear Disk (black profile); the
    channel resolution is 1.6 \kms. The red profile is the result of
    the Gaussian analysis.}    
  \label{2}
\end{figure}

\begin{table*}
\caption{Gaussian fits to the +50 \kms\ Cloud \nht\ profile.}
\begin{flushleft}
\begin{tabular}{lllllll}
\hline\hline\noalign{\smallskip}
$V$  & $T^*_{\rm A}$  & $\Delta V$  & $\tau_{o}$ &  $N$(\nht)  & Source \\
    (\kms)   &   (mK)   &   (\kms)  &    &  (cm$^{-2}$) & \\ 
\hline\noalign{\smallskip}
$-230$ & $-36 \pm 8$ &  7 & 0.19$\pm 0.05$  & $(5\pm 1) \times 10^{12}$ & High
Negative Velocity Gas (HNVG)  \\
$-217$ & $-41 \pm 7$ &  9 & 0.22$\pm 0.04$  & $(7\pm 1) \times 10^{12}$ & HNVG \\   
$-189$ & $-29 \pm 8$ &  7 & 0.15$\pm 0.04$  & $(4\pm 1) \times 10^{12}$ &  HNVG \\
$-136$ & $-39 \pm 8$ & 11 & 0.21$\pm 0.04$  & $(8\pm 2) \times 10^{12}$ & Expanding
Molecular Ring (EMR) (near side) \\
$-111$ & $-53 \pm 3$ & 51 & 0.29$\pm 0.02$  & $(55\pm 3) \times 10^{12}$ &  EMR
(near side) \\
$-48$  & $-40 \pm 6$ & 11 & 0.21$\pm 0.03$  & $(9\pm 1) \times 10^{12}$ &  3-kpc arm
\\
$-29$  & $-19 \pm 9$ &  7 & 0.10$\pm 0.04$  & $(2\pm 1) \times 10^{12}$ &  $-30$
\kms\ arm \\
$-3$   & $-45 \pm 39$ & 12 & 0.24$\pm 0.20$  & $(10\pm 9) \times 10^{12}$ &
Local/Sgr arm \\
$47$   & $266 \pm 29$ & 39 & ...  & $(1\pm 0.1) \times 10^{16}$
  & +50 \kms\ Cloud \tablefootmark{a} \\ 
$49$   & $-96 \pm 28$ & 18 & 0.23$\pm 0.07$  & $(15\pm 5) \times 10^{12}$
  & +50 \kms\ Cloud \tablefootmark{b} \\
$186$  & $28 \pm 8$  & 7 & ...   & ...  & ...  \\
 
\noalign{\smallskip}\hline\end{tabular}
\end{flushleft}
 $^a$ From RADEX analysis, see text \\
 $^b$ Self-absorption, see text \\ 
\end{table*}

The major emission feature in the +50 \kms\ profile, at 47 \kms\,
originates {\it in} the +50 \kms\ Cloud. The Gaussian analysis yielded
a $T^*_{\rm A} = 266$ mK and a line width of 39 \kms. Correcting
for the beam efficiency (0.89) we obtain a main beam brightness
temperature of 298 mK. In order to obtain an \nht\ column density for
the cloud, we have used the on-line version of RADEX
\footnote{http://var.sron.nl/radex/radex.php} (van der Tak et
al. \cite{van07}). Assuming a gas temperature of $T=80$ K and a gas
density of $n_{\rm H_2}=10^4$ cm$^{-3}$ (Walmsley et al. \cite{wal86};
Sandqvist et al. \cite{san08}) and a line width of 39 \kms, we get
$N$(\nht) $=1 \times 10^{16}$ cm$^{-2}$. This is comparable to the
value of $2 \times 10^{16}$ deduced by Mills \& Morris
(\cite{mil13}) from their observations of 14 higher \nht\ transitions
in this cloud, where they estimate that about 10\% of the \nht\ column
originates from a high temperature ($\approx 400$ K) cloud
component. It is, however, larger than the value of $3 \times
  10^{15}$ cm$^{-2}$ obtained by Herrnstein \& Ho (\cite{her05}) from
  their observations of the 23 GHz inversion line.
For the +50 \kms\ Cloud, our column density value then yields an
o-\nht\ abundance with respect to hydrogen,
$X$(\nht)$=N$(\nht)$/N$(\htwo), of $4 \times 10^{-8}$, if we assume a
molecular hydrogen column density of $N$(\htwo)$=2.4 \times 10^{23}$
cm$^{-2}$ (Lis \&\ Carlstrom \cite{lis94}), which was determined from
observations of a total dust column. Using instead the value of
$N$(\htwo)$=1.6 \times 10^{23}$ cm$^{-2}$, which was determined from
SEST observations of the \ciso\ (1-0) and (2-1) lines at this
position, (Karlsson et al. \cite{kar13}), we get an o-\nht\ abundance
of $6.3 \times 10^{-8}$.

The emission profile of the +50 \kms\ Cloud is significantly affected
by self-absorption, as can be seen in Fig. 1. Also, the Gaussian
analysis resulted in a clear self-absorption feature at 49 \kms\
(see Table 1) with an intensity of $-96$ mK. The corresponding
optical depth and column density for this feature were then
determined, using Eqs. (2) and (1), assuming that the self-absorbing
region of the cloud is on the near side of the +50 \kms\ Cloud and thus
absorbing both the background continuum of 210 mK and the
background line emission of 265 mK, a total of 475 mK. 

\begin{table*}
\caption{Gaussian fits to the CND SW \nht\ profile.}
\begin{flushleft}
\begin{tabular}{lllllll}
\hline\hline\noalign{\smallskip}
$V$  & $T^*_{\rm A}$  & $\Delta V$  & $\tau_{o}$ &  $N$(\nht)  & Source \\
    (\kms)   &   (mK)   &   (\kms)  &    &  (cm$^{-2}$) & \\ 
\hline\noalign{\smallskip}
$-220$ & $-43 \pm 8$ &  7 & 0.36$\pm 0.08$  & $(9\pm 2) \times 10^{12}$ &  HNVG \\
$-195$ & $-34 \pm 5$ & 17 & 0.28$\pm 0.05$  & $(17\pm 3) \times 10^{12}$ &  HNVG \\ 
$-134$ & $-49 \pm 12$ &  4 & 0.43$\pm 0.13$  & $(6\pm 2) \times 10^{12}$ &  EMR
(near side) \\
$-113$ & $-31 \pm 6$ & 15 & 0.25$\pm 0.05$  & $(14\pm 3) \times 10^{12}$ &  EMR
(near side) \\
$-85$  & $-44 \pm 8$ &  8 & 0.38$\pm 0.08$  & $(11\pm 2) \times 10^{12}$ &  CND \\
$-51$  & $-43 \pm 7$ &  9 & 0.37$\pm 0.07$  & $(13\pm 2) \times 10^{12}$ &  3-kpc
arm \\
$-22$  & $-19 \pm 8$ &  8 & 0.14$\pm 0.06$  & $(4\pm 2) \times 10^{12}$ &  $-30$
\kms\ arm \\
$3$    & $-44 \pm 11$ &  4 & 0.38$\pm 0.11$  & $(6\pm 2) \times 10^{12}$ &
Local/Sgr arm \\
$33$   & $139 \pm 47$ & 27 & ...   & $(4\pm 1) \times 10^{15}$
  & Molecular Belt \tablefootmark{a} \\ 
$34$   & $-64 \pm 50$ & 13 & 0.23$\pm 0.19$  & $(11\pm 9) \times 10^{12}$
  & Molecular Belt \tablefootmark{b} \\ 
$68$   & $46 \pm 6$  & 81 & ...   & $(7\pm 1) \times 10^{14}$
  & Shock region \tablefootmark{a} \\
$76$   & $-31 \pm 8$ & 9  & 0.19$\pm 0.05$ to 1.18$\pm 0.6$    &$((6\pm
 2) {\rm to} (40\pm 20)) \times 10^{12} $
  & ?? \tablefootmark{b} \\  
 
\noalign{\smallskip}\hline\end{tabular}
\end{flushleft}
 $^a$ From RADEX analysis, see text \\
 $^b$ Self-absorption, see text \\ 
\end{table*}

There are two overlapping emission features in the CND SW profile seen
in Fig. 2 and analysed in Table 2, namely at 33 and 68 \kms\ with
line widths of 27 and 81 \kms, 
respectively. In both of these features there are signs of
self-absorption at 34 and 76 \kms, respectively. The optical depths
of these self-absorptions will depend on the relative positions of the
components along the line of sight. For the 34 \kms\ self-absorption
we determine the optical depth, assuming that it and its emission
companion at 33 \kms\ are in front of the emission component at 76
\kms, and that the self-absorption is on the near side of the 33
\kms\ component. Thus the self-absorbing region absorbs the continuum
emission of 140 K and both the 33 \kms\ emission and the blue
overlapping wing of the 68 \kms\ emission feature, which amounts to
a total sum of $140+138+28=306$ mK, a value we use for $T_{\rm cont}$ 
in Eq. (2). For the 76 \kms\ self-absorption, we assume that the
self-absorbing region is on the near side of the 68 \kms\ emission
region and both lie either in front of, or behind, the 140 K continuum
region. Thus a range for the optical depth is obtained, using
either the red-shifted emission of the 68 \kms\ region at the
appropriate velocity of 76 \kms\, whose emission value is 45 mK, or
$140+45=185$ mK, the values of which we use for $T_{\rm cont}$ in
Eq. (2).

There are two major emission features in the CND SW profile and we
have performed a RADEX analysis of both. First we assume that the 33
\kms\ feature originates in the Molecular belt and we can thus apply
its physical properties to the analysis, namely a gas temperature of
80 K and density of $10^4$ cm$^{-3}$. The antenna temperature of 139
mK converts to a main beam temperature of 156 mK; the line width is
27 \kms. The best RADEX fit then yields an \nht\ column density of
$4 \times 10^{15}$ cm$^{-2}$.

\begin{figure}
  \resizebox{\hsize}{!}{\rotatebox{270}{\includegraphics{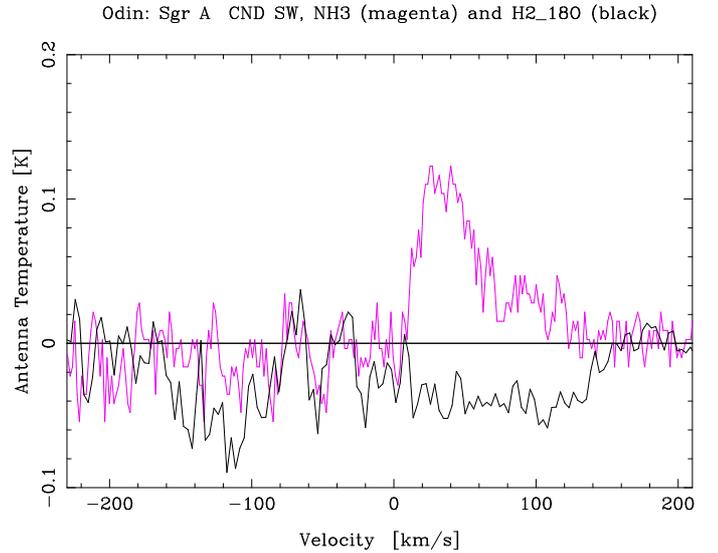}}}
  \caption{Comparison of the {\it Odin} observations of the 572 GHz
    \nht\ line (magenta) and the 548 GHz \htio\ line (black) towards
    the southwestern part of the \sgra\ Circumnuclear Disk; the
    channel resolutions are 1.6 and 3 \kms, respectively.}    
  \label{3}
\end{figure}

Perhaps the most interesting feature is the broad \nht\ emission feature
centered at 68 \kms\ with a line width of 81 \kms. {\it Odin} has
detected a comparable very broad \htio\ absorption line at the same
position (Karlsson et al. \cite{kar15}) - see Fig. 3. The antenna temperature of
this \nht\ feature is 46 mK, which implies a main beam temperature of 52
mK. Applying a RADEX study, and assuming a kinetic temperature of 200 K
and density of $4 \times 10^{4}$ cm$^{-3}$ (Requena-Torres et
al. \cite{req12}), we obtain a column density of o-NH$_3$ of $7
\times 10^{14}$ cm$^{-2}$. The \htwo\ column density for this region
is $4 \times 10^{22}$ cm$^{-2}$ (Karlsson et al. \cite{kar15}), which
would then yield an abundance ratio for o-\nht\ with respect to
\htwo\ of $X[\rm {NH_3}]=1.75 \times 10^{-8}$. Karlsson et al. (\cite{kar15})
obtained a strikingly high \hto\ abundance of $1.4 \times 10^{-6}$ in
this shock region.

\begin{table*}
\caption{\nht\ and \hto\ abundance comparisons}
\begin{flushleft}
\begin{tabular}{lllll}
\hline\hline\noalign{\smallskip}
Source & $N$(\nht)  & $N$(\htwo) & $X$(\nht) &  $X$(\hto)  \\
     &  (cm$^{-2}$) & (cm$^{-2}$)  & ($\times 10^{-9}$) &  ($\times 10^{-9}$) \\ 
\hline\noalign{\smallskip}
{\it \sgracomp\ Components} \\
+50 \kms\ Cloud & $(1\pm 0.1) \times 10^{16}$ & $1.6 \times 10^{23}$
\tablefootmark{a}   & 63 &  40 \tablefootmark{a} \\
+50 \kms\ Cloud red wing & ...  & ...  & ... &  $\ge 1000$  \tablefootmark{a} \\
Molecular Belt at CND SW & $(4\pm 1) \times 10^{15}$  &  <1.0$ \times
10^{23}$ \tablefootmark{b}  & >40  & ... \\ 
Shock region at CND SW & $(7\pm 1) \times 10^{14}$ &  $4 \times 10^{22}$
\tablefootmark{c} & 18   & 1400 \tablefootmark{c} \\
+20 \kms\ Cloud & ...  & $1 \times 10^{23}$ \tablefootmark{b} & ...  & 20
\tablefootmark{a} \\ 
+20 \kms\ Cloud red wing & ...  & ...  & ...  & $\ge 800$ \tablefootmark{a} \\
{\it Galactic Components}  \\
EMR & $(30\pm 2) \times 10^{12}$ \tablefootmark{d} & $2.5 \times 10^{21}$
\tablefootmark{e}  & 12 & ...  \\ 
3-kpc Arm & $(11\pm 2) \times 10^{12}$ \tablefootmark{d}  & $4.0 \times 10^{21}$
\tablefootmark{f}   & 3  & >3 \tablefootmark{f} \\
 $-30$ \kms\ Arm & $(3\pm 1) \times 10^{12}$ \tablefootmark{d} &   $1.9 \times 10^{21}$
 \tablefootmark{f} & 2 & 30 \tablefootmark{a} \\
Local/Sgr Arm & $(8\pm 5) \times 10^{12}$ \tablefootmark{d} &  $6.4 \times 10^{21}$
\tablefootmark{f}   & 1  & >3 \tablefootmark{f} \\

\noalign{\smallskip}\hline\end{tabular}
\end{flushleft}
 $^a$ Karlsson et al. (\cite{kar13});
 $^b$ value for the +20 \kms\ Cloud from Karlsson et al. (\cite{kar13});
 $^c$ Karlsson et al. (\cite{kar15});
 $^d$ mean of the respective values in Tables 1 and 2;
 $^e$ deduced from the $^{13}$CO($1-0$) profile in Fig. 1(a) by Bally
 et al. (\cite{bal88}) and the conversion factor given by Sofue (\cite{sof95});
 $^f$ Sandqvist et al. (\cite{san03}) 
\end{table*}

A summary of the derived \nht\ abundances for different components
in the \sgracomp\ and Galactic features is presented in Table 3,
together with \hto\ abundances, where available. 
It appears that the ortho-ammonia abundances of $(2-6) \times 10^{-8}$
observed by us in the \sgra\ molecular cloud regions  (the +50 \kms\
Cloud, as well as the Molecular Belt) can be accommodated in current
chemical models for dense gas clouds (Pineau des Forets et al. \cite{pin90}),
especially so if grain surface reactions and various subsequent
desorption processes are considered (Persson et al. \cite{per14}; dense cloud
part of their Fig. C2 ). Our detections of \nht\ at the much lower
abundance of $(1-3) \times 10^{-9}$ in spiral arm clouds are not so
easy to understand in terms of chemical models. 
Similarly low abundances have been derived by Persson et al. (\cite{per10},
\cite{per12}) from {\it Herschel} observations of spiral arm \nht\
absorption regions in the directions of W49N and G10.6$-$0.4 (W31C),
and by Wirstr\"om et al. (\cite{wir10}) from {\it Odin}
  observations in the direction of Sgr\,B2. The
  \nht\ abundances in diffuse or translucent clouds, estimated by Liszt
  et al. (\cite{lis06}) from 23 GHz inversion line absorption against compact
  extragalactic sources, are also similarly low. The low \htwo\ column
densities of the spiral arm clouds correspond to only a
few magnitudes of visual extinction \footnote{using
  $N$(\htwo)$/A_{\rm v} \approx0.94 \times 10^{21}$ cm$^{-2}$
mag$^{-1}$, Bohlin et al. (\cite{boh78})} and rather low  
cloud densities (so called translucent clouds).  However, it seems
that a chemical model combining gas-phase and grain surface reactions
can do the job at a temperature of 30-50 K and a density of $\approx10^3$
cm$^{-3}$, but only if the relevant species really are desorbed as a result
of exothermic surface reactions (Persson et al. \cite{per14}; their Fig.C3,
and the translucent gas part of Fig. C2). A remaining concern may be
that the simultaneous model abundances of NH and NH$_2$ are not
consistent with those observed by Persson et al. (\cite{per10}, 2012).  

As discussed in some detail by Karlsson et al. (\cite{kar13}), the gas-phase
water abundances  determined for the Sgr A +50 \kms\ and +20 \kms\
molecular clouds are considerably enhanced compared to the situation
in cold cloud cores where the water mainly resides as ice on the cold
grain surfaces. Similarly high or even more enhanced water abundances
have been estimated in lower density spiral arm clouds observed in
absorption against Sgr A by Karlsson et al. (\cite{kar13}), and against Sgr B2
using {\it Odin} by Wirström et al. (\cite{wir10}) and using {\it
  Herschel} Space Observatory also by Lis et al. (\cite{lis10}). The
elevated gas phase water abundances definitely require desorption of
water ice, most likely handled by PDR modeling including grain surface reactions
(cf. Hollenbach et al. \cite{hol09}) The release of water molecules formed on
colder grain surfaces may also explain the ortho-to-para \hto\ abundance
ratio lower than 3 very likely observed in the spiral arms against Sgr
B2 (Lis et al. \cite{lis10}). 

The very high gas-phase water abundance determined for the shock
region at CND SW by Karlsson et al. (\cite{kar15}) is similar to that found in
the red-ward high-velocity wings of the Sgr A molecular clouds, and
likely results from shock heating causing release of pre-existing
grain surface water, possibly combined with high temperature shock
chemistry. (cf. discussion in Karlsson et al. \cite{kar13}). The very large
velocity width (80 \kms) of the \nht\ emission associated
with the shock region at the CND SW (Fig. 2 and Table 2) may suggest
a formation/desorption scenario similar to that of gas-phase \hto\ in
shocks/outflows, where the actual ammonia abundance is achievable in
chemical models for quiescent molecular cloud (as discussed previously).   

\section{Conclusions}

We have successfully observed \nht\ emission from the \sgra\ +50 \kms\
Cloud and the CND using the {\it Odin} satellite.
The very large velocity width (80 \kms) of the \nht\ emission
associated with the shock region in the southwestern part of the CND
may suggest a formation/desorption scenario similar to that of
gas-phase \hto\ in shocks/outflows. In addition, clear \nht\
absorption has also been observed in the spiral arm features along the
line of sight to the Galactic  Centre.

The high quality {\it Odin} \nht\ spectra, observed in 2015 and 2016, presented
in this short paper, together with the  \hto\ and \htio\ spectra of similarly
high quality observed by {\it Odin} towards Comet C/2014 Q2 (Lovejoy) in
January - February 2015 (Biver et al. \cite{biv16}) were obtained 14-15 years
after the launch of the {\it Odin} satellite. This demonstrates to our
delight (and also some surprise) that it is possible to build a
comparatively cheap, but still rather complicated, satellite
observatory (with five tunable heterodyne receivers and a mechanical
cooling machine) which can remain in high quality operation for 15
years or more. We should here note that the {\it Odin} satellite
observatory since many years has been operated practically full-time in
global monitoring of the terrestrial atmosphere (the {\it Odin}
aeronomy mode).

\end{document}